\documentclass[twocolumn, prl]{revtex4-1}
\usepackage[utf8]{inputenc}
\usepackage{amsmath}
\usepackage{amssymb}
\usepackage{graphicx}
\usepackage{xcolor}
 \usepackage{units}
\usepackage{hyperref}
 \usepackage[nottoc,numbib]{tocbibind}
 \newcommand{\figref}{Figure \ref}
 \newcommand{\Eqref}{Eq. \ref}
 
 \newcommand{\xdownarrow}[1]{%
  {\left\downarrow\vbox to #1{}\right.\kern-\nulldelimiterspace}
}
\begin{document}

\title{Exploring Non-Abelian Geometric Phases in Spin-1 Ultracold Atoms}
\author{Bharath H. M.$^{1, 2}$, Matthew Boguslawski$^{1}$, Maryrose Barrios$^{1}$, Lin Xin$^{1}$ and M. S. Chapman$^{*1}$ }

\date{\today}
\affiliation{$^{1}$School of Physics, Georgia Institute of Technology.}
\affiliation{$^2$Ludwig-Maximilians-Univesit\"at M\"unchen.}

\begin{abstract}
Non-Abelian and non-adiabatic variants of Berry's geometric phase have been pivotal in the recent advances in fault tolerant quantum computation gates, while Berry's phase itself is at the heart of the study of topological phases of matter.  The geometrical and topological properties of the phase space of spin$-1$ quantum states is richer than that of spin$-1/2$ quantum states and is relatively unexplored. For instance, the spin vector of a spin-1 system, unlike that of a spin$-1/2$ system, can lie anywhere on or inside the Bloch sphere representing the phase space. Recently, a generalization of Berry's phase that encapsulates the topology of spin-1 quantum states has been formulated in \textit{J. Math. Phys.}, 59(6), 062105. This geometric phase includes loops that go inside the Bloch sphere and is carried by the tensor of spin fluctuations, unlike Berry's phase which is carried by the global phase of the quantum state. Furthermore, due to a mathematical singularity at the center of the Bloch sphere,  the class of loops that pass through the center are called singular loops and are significant because their geometric phase is non-Abelian. In contrast with Berry's phase for spin$-1/2$ systems, whose properties come from the topology of a sphere, the properties of singular loop geometric phases come from the topology of the \textit{real projective plane} $\mathbb{RP}^2$, which is more non-trivial. Here we use coherent control of ultracold $^{87}$Rb atoms in an optical trap to experimentally explore this geometric phase for singular loops in a spin-1 quantum system.  
\end{abstract}
\maketitle

Berry's geometric phase is a manifestation of the geometrical and topological properties of the state space (or parameter space) of the physical system, with no regards to the dynamics. Although, the condition of \textit{adiabaticity} in the first example of Berry's phase \cite{Berry.1984} speaks of the dynamics, it was later established that this condition is dispensable \cite{PhysRevLett.58.1593} and that more generally, geometric phase is purely a manifestation of the geometry of the underlying space \cite{PhysRevLett.51.2167}, independent of the dynamics, and therefore it can be formulated as a kinematic property of paths in the underlying space \cite{Mukunda-1}. For instance, fully magnetized spin states live on a sphere, known as the Bloch sphere. Consequently, Berry's phase for such states is a manifestation of the geometry and topology of a sphere. In contrast, the zero magnetization states of an integer spin system do not live on a sphere --- they live on the \textit{real projective plane} ($\mathbb{RP}^2$). It is the space of all diameters of a sphere and its topology is richer than that of a sphere. It is also the configuration space of nematic liquid crystals. As a consequence, the Berry phase of the zero magnetization states is a manifestation of the topology of $\mathbb{RP}^2$ and has been studied exclusively\cite{0305-4470-27-12-007}. This is an example of the richness of the topological properties of the space of spin$-1$ quantum states. Recent theoretical work \cite{Theory_Paper} has exposed other interesting features of the space of spin-1 quantum states, developing a non-Abelian geometric phase for loops inside the Bloch sphere, that also derives its properties from the topology of $\mathbb{RP}^2$. 

Berry phase originating from the geometry of a sphere has been widely explored. Owing to its geometrical origin, Berry phase for spin$-1/2$ systems has been shown to be robust to dynamical fluctuations. As a result, robust phase gates can be constructed out of this Berry phase and are known as \textit{holonomic gates} \cite{2016GeometricQC,  Kitaev2003, HolonomicQC1999, SJOQVIST201665,PhysRevA.92.052302}. Adiabatic holonomic gates in two-level systems have been demonstrated using nitrogen vacancy centers \cite{2016NaPho10184Y}, and solid state qubits \cite{Leek1889}. 
Non-Abelian, nonadiabatic holonomic gates have been demonstrated using microwave induced control in NV centers \cite{2014Natur51472Z, ArroyoCamejo2014RoomTH, polarised_microwaves} and transmon systems \cite{2013Natur496482A}. More recently, optically controlled holonomic gates have been implemented in NV centers \cite{OpticalHolNatPh, 2017arXiv170500654Z}, ion traps \cite{PhysRevA.87.052307} and NMR systems \cite{PhysRevLett.110.190501}. However, the geometric and topological properties exclusive to the space of spin-1 quantum states remains experimentally unexplored.

Here, we report on the first experimental observation, using ultracold $^{87}$Rb atoms, of the non-Abelian variant of geometric phase unique to spin$-1$ and higher systems introduced in \cite{Theory_Paper}. Apart from deriving its properties from the topology of $\mathbb{RP}^2$, this geometric phase is richer than Berry's phase in many other ways: it is defined for all loops on or inside the Bloch sphere and it is carried not by the overall phase, but by the tensor of spin fluctuations. As we detail in this paper, the latter, represented by a 3D ellipsoid, provides a useful geometric perspective on the properties of spin$-1$ quantum states. Unlike a global phase, the spin fluctuations are accessible for observables of the system and can be measured without requiring an interference.   

\begin{figure*}[ht!]
\includegraphics[scale=0.64]{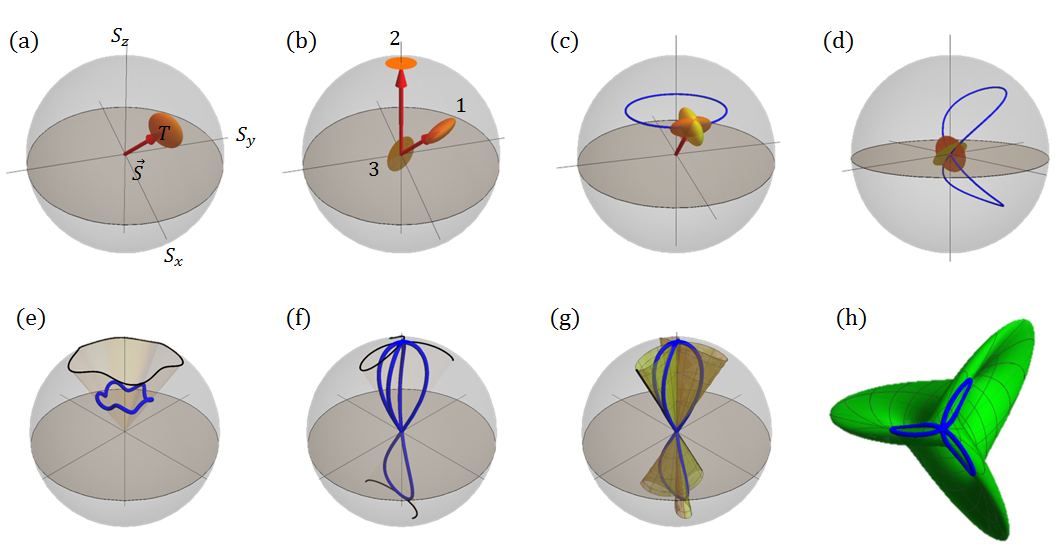}
\caption{\textbf{Theory of singular loop geometric phases:} (a) and (b) show a geometric representation of spin-1 quantum states. (a) shows that the spin vector ($\vec{S}$) and the spin fluctuation tensor ($T$) can together be represented by a point inside the Bloch sphere surrounded by an ellipsoid. This pair of the vector and the ellipsoid uniquely represent a spin-1 quantum state up to an overall phase.  (b) illustrates that the lengths of the ellipsoid's axes are constrained by the length of the spin vector. Explicitly, they are given by $\sqrt{1-|\vec{S}|^2}$ and $\sqrt{\frac{1\pm \sqrt{1-|\vec{S}|^2}}{2}}$. For the three examples labelled $1,2$ \& $3$, the spin vectors $\vec{S}_{1,2,3}$ satisfy $0<|\vec{S}_1|<1, |\vec{S}_2|=1$ and $|\vec{S}_3|=0$. The ellipsoid degenerates to a disk for the last two cases. (c) and (d) show the geometric phases carried by the ellipsoid when it is parallel transported along a non-singular and a singular loop inside the Bloch sphere respectively. In either of these cases, the final orientation of the ellipsoid is different from the initial orientation, due to an $SO(3)$ geometric phase. For singular loops, this geometric phase is non-Abelian. (e) and (f) contrast non-singular and singular loops under a radial projection. The former has a continuous projection and a well defined solid angle, while the latter doesn't. This problem is resolved by defining a generlized solid angle for singular loops using a diametric projection, as illustrtaed by (g) and (h). (g) shows a surface obtained by sweeping a diameter along the loop. The solid angle enclosed by this surface is the generalized solid angle of the singular loop. This surface is indeed a loop in the space of diameters of a sphere, i.e., in a real projective plane ($\mathbb{RP}^2$). (h) shows a Boy's surface, a representation of the real projective plane, together with the loop projected on it. The generalized solid angle is equal to the holonomy of this loop.}\label{FIG1}
\end{figure*}

We begin by briefly summarizing the theory of singular loop geometric phases. The quantum state of a spin$-1/2$ system is uniquely represented by a point on the Bloch sphere whose coordinates are given by the expectation values of the spin operators $S_x, S_y$ and $S_z$. Spin-1 (and higher) quantum states differ in two ways --- first, the expectation value of the spin vector, $\vec{S}=(\langle S_x\rangle, \langle S_y\rangle, \langle S_z \rangle)^T$ (here, $\langle \cdot \rangle$ represents the expectation value) is not confined to the surface of the Bloch sphere; it could be anywhere on or \textit{inside} the Bloch sphere. And second, a quantum state is not uniquely represented by its spin vector; there can be several different quantum states which share the same spin vector. For spin-1 systems, this ambiguity is resolved by considering the quantum fluctuations of the spin vector, which, geometrically, is an ellipsoid surrounding the head of the spin vector (\figref{FIG1} (a)). The ellipsoid represents a rank two tensor ($T$), whose components are the expectation values of the quadratic spin operators $T_{ij}=\frac{1}{2}\langle \{S_i, S_j\}\rangle-\langle S_i \rangle \langle S_j \rangle$. The pair $(\vec{S}, T)$ uniquely represents a spin-1 quantum state up to an overall phase (Supplementary Information). \figref{FIG1} (b) shows three examples.

Geometric phase arises in this system when the ellipsoid is parallel transported along a closed loop inside the Bloch sphere (\figref{FIG1}(c, d)).  As a result of the parallel transport, the ellipsoid returns in a different orientation which can be described by a 3D rotation, represented by a $3\times 3$ matrix. This rotation matrix ($R$), a member of the $SO(3)$ group, is the geometric phase of the loop. This geometric phase is an operator, unlike Berry's phase which is a complex scalar, and is therefore more similar to Wilczek-Zee phase \cite{PhysRevLett.52.2111} and Uhlmann phase \cite{Uhlmann.1986}, both of which are unitary matrices. This can be measured easily in the components of the spin fluctuation tensor, specifically, the component $T_{ij} $ changes to $R_{il}T_{lk}R_{jk}$ after the parallel transport. 
 \begin{figure*}[ht!]
\includegraphics[scale=0.62]{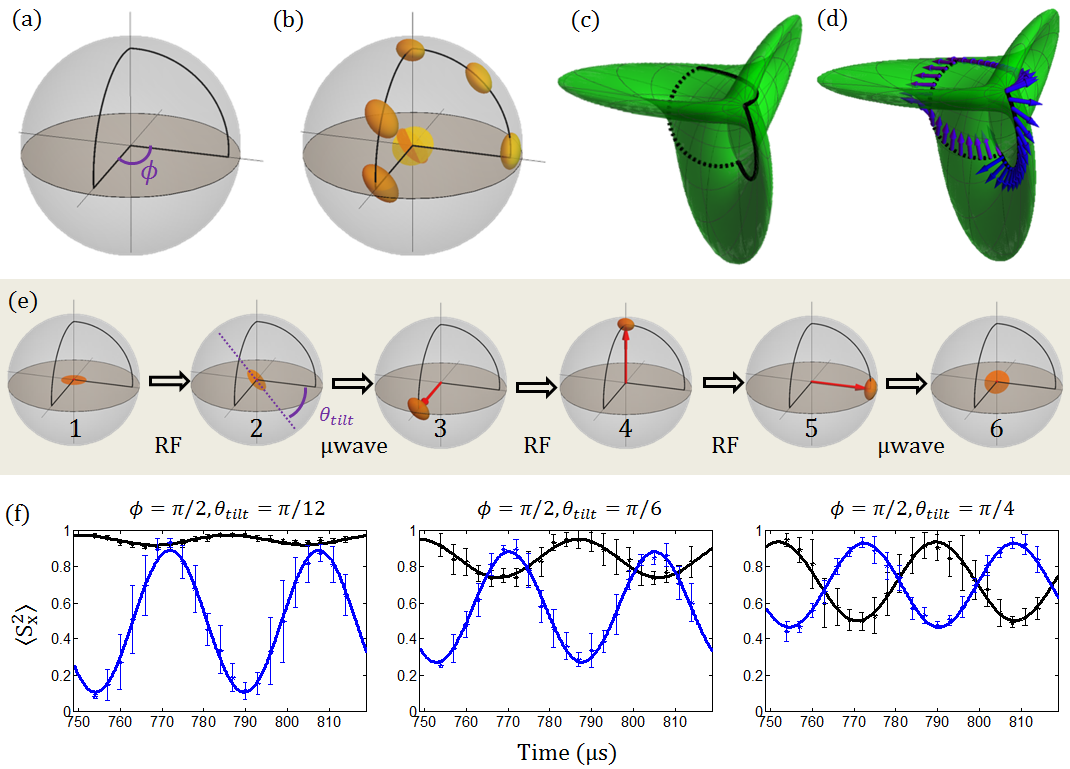}
\caption{\textbf{Experimental sequence:} (a) shows the singular loop that we choose to implement experimentally. It starts and ends at the center of the Bloch sphere. (b) shows how an ellipsoid is parallel transported along this loop. In particular, it starts out as a disk at the center and returns in a different orientation, rotated according to the geometric phase of the loop which is an $SO(3)$ operator. The generalized solid angle of this loop is given by the holonomy of its projection on a Boy's surface (real projective plane). (c) shows this projection. It is an open path and its holonomy is defined by closing it with a geodesic \cite{PhysRevLett.60.2339, Pancharatnam1956}, shown by the dashed curve. (d) illustrates the holonomy of this path, i.e., the angle of rotation of a unit tangent vector to the Boy's surface, after it is parallel transported along this loop. The experimental sequence of transporting the ellipsoid along this loop inside the Bloch sphere is illustrated in (e). Starting from a flat disk, an arbitrary tilt is induced using an RF pulse. Following, the loop is induced using microwave pulses for each the radial segment and RF pulses for each curved segment of the loop. Finally, in order to observe the geometric phase, we measure $S_x^2$ as the tilted disk spins about the $z$ axis at the Larmor rate. (f) shows the oscillation of $S_x^2$ without (black) and with (blue) the transport along the loop. The geometric phase is encoded in the phase shift and the amplitude shift between the black and the blue datasets. Each data point shown above is an average of ten shots and the error bars represent the standard deviation.   }\label{FIG2} 
 \end{figure*}

The parallel transport of the ellipsoid has a deep geometrical significance to the abstract space of quantum states. The Fubini-Study metric, also known as the ``quantum angle" characterizes the geometry of the space of quantum states \cite{Geom.of.quant.states}. Among the infinitely many ways of transporting the ellipsoid along a loop inside the Bloch sphere, the parallel transport is a special one; it minimizes the Fubini-Study length of the resulting path in the space of quantum states\cite{Uhlmann1995, Uhlmann2000}. Given any initial orientation of the ellipsoid, this path in the space of quantum states is well defined and unique. In the special case where the loop starts at the center of the Bloch sphere, a parallel transport is well defined and unique if the initial orientation of the disk is tangential to the loop.

Geometrical interpretation of this geometric phase, particularly for singular loops, needs an extended notion of solid angles introduced in \cite{Theory_Paper} as \textit{generalized solid angles}. For a non-singular loop, the geometric phase is a rotation about the spin vector by an angle equal to the solid angle of the loop (\figref{FIG1}(e)) and is therefore Abelian. This is because the parallel transport of the ellipsoid inside the Bloch sphere along a non-singular loop is reminiscent of the parallel transport of a tangent vector to a sphere. The solid angle of a non-singular loop is the angle of the cone obtained by sweeping a radius along the loop (\figref{FIG1}(e)), which produces a \textit{radial projection} of the loop. For the case of singular loops, this geometric notion of solid angles is not well defined, as illustrated in \figref{FIG1}(f). The radial projection is discontinuous and therefore, such loops require a generalization of the notion of solid angles. 

The key idea behind generalized solid angles is to use \textit{diametric} projections, instead of radial projections. The discontinuous jumps in a radial projection of singular loops are always diametrically opposite (\figref{FIG1}(f)) and therefore, sweeping a diameter along the loop generates a continuous cone with a well defined angle (\figref{FIG1}(g)). This angle is equal to the standard solid angle for non-singular loops and is a convenient generalization to singular loops.

While the standard solid angle is the integrated curvature or \textit{holonomy} of a loop on a sphere, the generalized solid angle is the holonomy of a loop on a real projective plane(Supplementary Information). $\mathbb{RP}^2$ can be represented by a self intersecting surface, known as Boy's surface \cite{Modelsofrp2} (\figref{FIG1}(h)). The cone generated by sweeping a diameter along a loop inside the Bloch sphere represents a path in the real projective plane. Thus, using a diametric projection, a loop inside the Bloch sphere is projected to the real projective plane, and the holonomy of the projected path is defined as the generalized solid angle of the loop inside the Bloch sphere \cite{Theory_Paper}. If the projected path on $\mathbb{RP}^2$ is open, its generalized solid angle is defined after closing it with a geodesic ( \figref{FIG2}(c) and Supplementary Information, section IV. C.). The diametric projection  also equips us with a concise way of determining the geometric phase operator $R$. The two endpoints of the diameter trace out a pair of congruent loops on the surface of the Bloch sphere, which we may parametrize in time as $+\hat{n}(t)$ and $-\hat{n}(t)$ respectively. The geometric phase is then given by 
\begin{equation}
R= \mathcal{T}\text{exp}\left\{\int (\dot{\hat{n}}\hat{n}^T - \hat{n}\dot{\hat{n}}^T)dt\right\}
\end{equation}
Here, $\mathcal{T}$ refers to the time ordering operator and the integral is evaluated through the loop.

We now turn to the experimental measurements. The experiments are performed using ultracold $^{87}$Rb atoms confined in an optical dipole trap (Supplementary Information). The spin-1 quantum system is provided by the $F=1$ hyperfine level of the electronic ground state of the atom. The atoms are initialized in the $m_F =0$ state, which is a spin state located at the origin of the Bloch sphere whose fluctuations are a planar disk in the $x-y$ plane.  From this starting point, any path within the Bloch sphere can be induced by a combination of rotating (rf) magnetic field pulses and microwave $2 \pi$ pulse connecting the $F,M_F = 1,0 \rightarrow 2,0$ states.  The former generates the familiar Rabi rotation of the spin, and the latter realizes a quadrupole operator that changes the spin length \cite{Hoang23082016, 2012NatPhys305H}. The final state of the system is determined by measuring the populations in $m_F=0, \pm 1$  using a Stern-Gerlach separation of the cloud followed by a fluorescence imaging of the atoms  \cite{Hoang23082016}. This provides a direct measurement of $\langle S_z\rangle$ and $\langle S_z^2\rangle$. The transverse components of the spin length and moments, e.g. $\langle S_x^2\rangle$, are measured using a $\pi/2$ rf pulse preceding the Stern-Gerlach separation.

To investigate the unique aspects of the geometric phase considered here, we use the class of loops shown in (\figref{FIG2}(a)). These are singular loops that start and end at the center (\figref{FIG2}(c)). These loops capture the distinguishing features of this geometric phase, and they are also convenient to realize experimentally. The experimental sequence is shown in 
\figref{FIG2}(e)). Starting from the initial state, an initial rf pulse is used to tilt the flat disk to the desired angle, $\theta_{tilt}$. We then induce a transport along the loop  using a sequence of microwave and rf pulses. In a frame rotating at the Larmor frequency, a resonant rf field is a constant field while the microwave fields are insensitive to the Larmor rotation. Therefore, the pulse sequence shown in \figref{FIG2}(e) effectively induces the loop in the rotating frame.

\begin{figure}[ht!]
\includegraphics[scale=0.70]{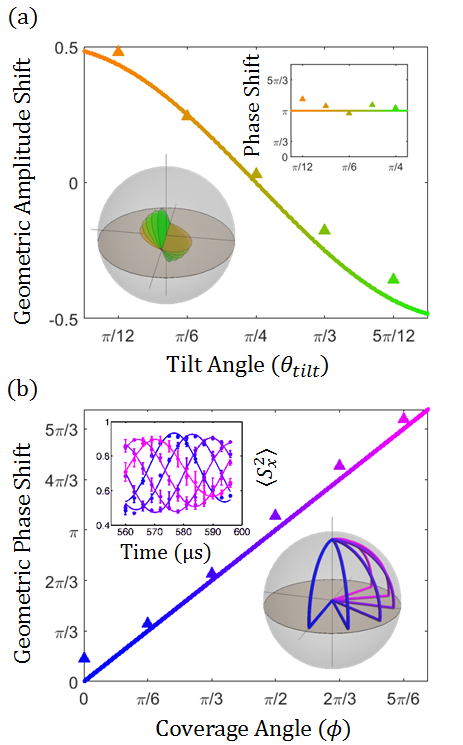}
\caption{\textbf{Geometric amplitude shift and phase shift:} (a) shows a comparison with theory of the experimentally observed geometric amplitude shifts. The theoretical value of this amplitude shift is $\frac{1}{2}\cos 2\theta_{tilt}$ (continuous curve). The triangular markers show the experimentally observed amplitude shifts for different tilt angles. The inset shows the geometric phase shifts for these five tilt angles and the continuous line shows the corresponding theoretical value, i.e., $\pi$. The bottom inset shows the disks (magnified) at the starting point with different tilt angles used in the experiment. (b) shows the geometric phase shift for different values of the coverage angle ($\phi$). The continuous line shows the theoretical geometric phase shift, i.e., $2\phi$. The upper inset shows the corresponding Larmor oscillation of $\langle S_x^2\rangle$. The lower inset shows the loops corresponding to the different values of $\phi$ used in the experiment. The error bars corresponding to the above data points  (i.e., fit parameters) range between $0.012$ and $0.03$, smaller than the markers and therefore are not displayed.}\label{FIG3} 
\end{figure}

For the loops shown in (\figref{FIG2}(a)), the generalized solid angle is $\phi$, and the corresponding geometric phase is $R=R_z(\phi)R_x(-\phi)$ (Supplementary Information). For the initial state, the spin fluctuation tensor is a disk at the center of the Bloch sphere intersecting the $x-y$ plane along the $x$-axis and making an angle $\theta_{tilt}$. When this disk is parallel transported along the indicated loop, the geometric phase $R$ manifests as a different final orientation of the disk, which now has an angle $\theta_{tilt}' = \phi+\theta_{tilt}$ with the $x-y$ plane and intersects it along the rotated axis $\hat{x'} = \hat{x}\cos \phi + \hat{y} \sin \phi$. 

Our experiments are done under a constant Larmor precession about the $z$-axis. Therefore, as a tilted disk at the center spins about the $z$-axis, $\langle S_x^2\rangle$ and $\langle S_y^2\rangle$ both oscillate at twice the Larmor frequency ($\omega_L$). In particular, if a disk is tilted by $\theta_{tilt}$ and intersects the $x-y$ plane along the $x$-axis at $t=0$, then $\langle S_x^2(t) \rangle = 1-\sin^2\theta_{tilt}\sin^2(\omega_Lt)$ and $\langle S_y^2(t) \rangle = 1-\sin^2\theta_{tilt}\cos^2(\omega_Lt)$. If it is parallel transported along the given loop in a frame rotating at the Larmor rate, the accumulated geometric phase can be observed by measuring $\langle S_x^2(t)\rangle$. It is straightforward to see that, after the parallel transport, $\langle S_x^2(t)\rangle = 1-\sin^2\theta_{tilt}'\sin^2(\omega_Lt+\phi)$. That is, the geometric phase can be observed as a phase shift as well as an amplitude shift of the oscillation of $\langle S_x^2(t)\rangle$. The geometric phase shift would be $2\phi$ and the amplitude shift would be $\sin^2\theta_{tilt}-\sin^2\theta_{tilt}'$.

We have measured both the geometric phase shifts and amplitude shifts for a range of angles 
 $(\theta_{tilt}, \phi)$ as shown in \figref{FIG3}. In \figref{FIG3}(a), we investigate loops with a fixed angle $\phi=\frac{\pi}{2}$ for different initial orientations of the disk, $\theta_{tilt}$ in order to demonstrate a nontrivial amplitude shift. The geometric phase of this loop is $R=R_z(\pi/2)R_x(-\pi/2)$ and the solid angle is $\pi/2$. The theoretical phase shift in the oscillation of $\langle S_x^2\rangle$ is $2\phi =\pi$ for each of the initial orientations of the disk,  and the experimental values are in good agreement as seen in the inset of \figref{FIG3}(a). The theoretical amplitude shift depends on the initial disk orientation --- it is $\frac{1}{2}\cos2\theta_{tilt}$. As can be seen in \figref{FIG3}(a), the observed amplitude shift is in excellent agreement with the theory. Data sets with explicit sinusoidal fits showing the phase shift and amplitude shift corresponding to three of the different initial orientations are shown in \figref{FIG2}(f).

In \figref{FIG3}(b), we demonstrate the dependence of the phase shift to the generalized solid angle of the loop.  For these measurements, the starting disk orientation is $\theta_{tilt}=\frac{\pi}{4}$ and the range of loops investigated is shown in the inset to the figure. The measured phase shifts show excellent agreement with the theoretical phase shift  in the oscillation of $\langle S_x^2\rangle$, which is $2\phi$. 

We note for the measurments in \figref{FIG3}(b), it is necessary to compare the results with reference loops with no geometric phase in order to isolate the geometric phase from the dynamical phase.  In the rotating frame, the transport induced by the rf pulse is naturally a parallel transport; i.e., the rf field evolves the system along the path of least Fubini-Study length \cite{2016GeometricQC}. However, this is not true for the microwave pulses; the transport along the straight segments is not parallel and the system is taken along a path of non-minimal Fubini-Study length (Supplementary Information). Consequently, some dynamical phase is accumulated during this transport that needs to be measured in order to isolate the geometric phase. To accomplish this, we take two data sets each measuring the oscillation of $\langle S_x^2\rangle$ --- one after transporting the disk along the loop and another after transporting the disk radially outward and then back inward, for which there is no geometric contribution. A comparison of these two data sets allows determination of the geometric phase shift and amplitude shift of the induced loops as shown in \figref{FIG2}(f). 

Unlike Abelian geometric phases, this geometric phase manifests in terms of two observable parameters --- the phase shift and the amplitude shift, both of which we have demonstrated. Using a set of loops, all of them based at the center, we have shown the variation of the amplitude shift at constant phase shift in one dataset (\figref{FIG3}(a)) and the variation of the phase shift at constant amplitude shift in the other (\figref{FIG3}(b)). In particular, we have shown experimentally that these two parameters vary independently, implying that the group of geometric phase operators is more complicated than the Abelian group of all rotations about a fixed axis --- it has more than one parameter. There are no two parameter Abelian subgroups in the group of all rotations, i.e., SO(3). Therefore, our experimental data can not be fit into an Abelian geometric phase and thus we have demonstrated the non-Abelian nature of this geometric phase, which occurs only for singular loops \cite{Theory_Paper}. 

We have demonstrated that as a result of their rich geometrical and topological properties,  spin$-1$ quantum states possess a unique non-Abelian geometric phase that is carried by the spin fluctuation tensor. The perspective we have taken here on spin$-1$ quantum states --- representing it as a spin vector together with an ellipsoid of spin fluctuation tensor has been crucial in revealing this distinctive geometric phase. We believe that this picture of spin$-1$ quantum states, which can be applied to any three level system, will expose further interesting geometrical and topological properties and is in general a useful way of characterizing quantum systems with a spin higher than $1/2$.  

A natural succession of  our experiment is to prepare 1D and 2D spin textures in spatially extended spin-1 systems. Spin textures have been extensively studied for spin$-1/2$ systems for instance, Skyrmions, Domain walls and N\'{e}el walls, and are characterized by their Berry phase. Spin-1 quantum states add a new feature to these textures --- the spin vector can be inside the Bloch sphere, allowing for a larger class of textures, whose topological properties are unexplored. The techniques demonstrated in this paper are a step towards experimentally studying such spin textures, and we believe that our geometric phase will play a role in understanding their properties.

Distinct from Berry's phase, our geometric phase is carried by the spin fluctuation tensor, which is a rank$-2$ tensor.  We hope that this idea will find applications in studying topological phase transitions \cite{2016TopologicalPhasesWitten} in systems with a complex tensor order \cite{PhysRevB.97.064504}. The order parameter in such systems is also a rank-2 tensor, and transforms similar to the spin fluctuation tensor under $SO(3)$ rotations. 
 
\section*{Acknowledgments}
We thank John Etnyre, T. A. B. Kennedy, Carlos S\'{a} de Melo and Shubhayu Chatterjee for fruitful discussions and insights. We also acknowledge support from the National Science Foundation, grant no. NSF PHYS-1506294.
\bibliography{GeometricPhasePaper}{}
\begin{figure*}
\begin{center}
    \Large \textbf{Supplementary Information}
\end{center}
\end{figure*}
\newpage

 In this document, we fill in the technical details of the four ideas discussed in the paper that are pivotal to our results:
\begin{itemize}
\item A spin-1 quantum state (excluding the overall phase) is uniquely represented by a point inside the Bloch sphere surrounded by an ellipsoid.
\item When this ellipsoid is parallel transported along a closed loop inside the Bloch sphere, it picks up an $SO(3)$ geometric phase.
\item There is a definition of geometric phases in general, that is completely independent of the system's dynamics.
\item The notion of solid angles enclosed by a loop on the Bloch sphere can be generalized to loops inside the Bloch sphere, including those that pass through the center.
\end{itemize}
In sections I, III, II and IV, we fill in the technical details of the above four ideas in that order. In section V, we briefly summarize the experimental control operations.
\section{Geometric coordinates for spin-1 quantum states}
A spin-1 quantum state is a three dimensional complex vector. Ignoring the overall phase, a normalized state vector has four independent parameters and therefore, the space of spin-1 quantum states is a four dimensional manifold. As mentioned in the main text, points in this manifold can be represented by the pair $(\vec{S}, T)$, of the spin vector and an ellipsoid, or, by an unordered pair of points on the Bloch sphere, known as Majorana constellation \cite{Geom.of.quant.states}. The latter comes from the observation that the symmetric (i.e., triplet) subspace of a pair of spin-1/2 systems is homemorphic to a spin-1 system. 

\begin{figure*}
\includegraphics[scale=0.6]{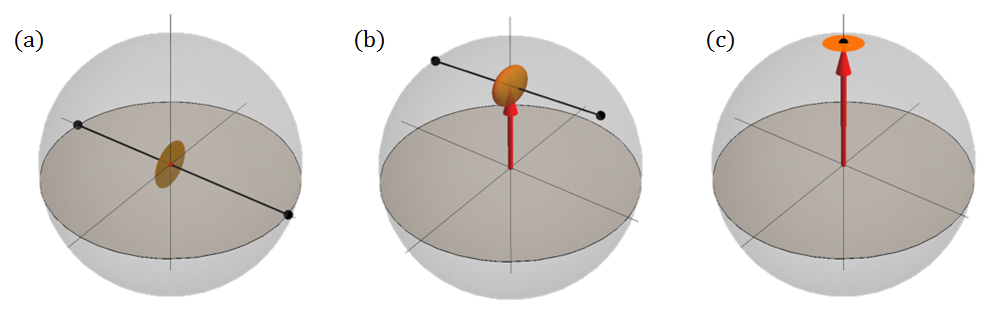}
\caption{\textbf{Majorana constellation:} Three example states represented by the pair $(\vec{S}, T)$ and by a pair of points (endpoints of the chord) on the Bloch spheres. }\label{FIG1S}
\end{figure*}

A pair of points on the Bloch sphere is represented by a pair of unit vectors  $(\hat{r}_1, \hat{r}_2)$. The corresponding spin vector is given by $\vec{S}=\frac{\hat{r}_1+\hat{r}_2}{2}$. It is straightforward to check that the ellipsoid of quantum fluctuations is oriented such that it's axes are parallel to $\hat{r}_1+\hat{r}_2$, $\hat{r}_1-\hat{r}_2$ and $\hat{r}_1\times \hat{r}_2$. The smaller of the axes normal to the spin vector is parallel to $\hat{r}_1-\hat{r}_2$ and we denote the corresponding unit vector by $\hat{u}=\frac{\hat{r}_1-\hat{r}_2}{|\hat{r}_1-\hat{r}_2|}$. While $(\hat{r}_1, \hat{r}_2)$ can be considered as geometric coordinates for a spin-1 state, an equivalent set of coordinates are $(\vec{S}, \hat{u})$. Note that $(\vec{S}, \hat{u})$ and $(\vec{S}, -\hat{u})$ represent the same state and therefore, we write $(\vec{S}, \pm \hat{u})$ (see \figref{FIG1S}). 

\section{Defining geometric phases using metrics}
Geometric phases are carried by the system's gauge variables. For instance, in Berry's phase of a spin-1/2 system, the overall phase of the the quantum state is the gauge variable. A point on the Bloch sphere does not completely specify the full quantum state vector; one has to append the overall phase, i.e., the gauge variable. Consequently, given a loop on the Bloch sphere, there are several ways of tuning the control parameters so as to transport a system along the loop. They all would induce the same loop, but differ in the profile of the overall phase along the loop. Of these, there is a special one, which corresponds to the parallel transport of the overall phase along the loop. Geometric phase of a loop is the mismatch between the initial and final overall phase values of the parallel transport. At the heart of this definition is the rule of parallel transport --- what does it mean to parallel transport the overall phase? One way to define parallel transports is to use a structure called a connection form, which we do not elaborate here.

In general, the various ways of tuning the control parameters that all induce the same given loop, differ not only in the profile of the gauge variable, but also in the distance traversed  in the full Hilbert space (including the gauge coordinate). Quite intriguingly, in all the well known examples of geometricphases, when the gauge variable is parallel transported, the system traverses the \textit{least} possible distance in the Hilbert space \cite{Uhlmann1995, PATI1991105}.  This prompts a more general definition of parallel transport --- to parallel transport a system is to minimize the distance traversed. If we tune the control parameters such that not only the given loop is induced, but also, the system travels the least distance in the full Hilbert space, then we have parallel transported the system.  This holds for all examples of parallel transport. Indeed, it is intuitive that when a state is being parallel transported on the Bloch sphere, we carefully avoid any ``unnecessary" changes to the overall phase.  This is consistent with the idea of minimizing the total distance traversed in the Hilbert space, because changes in the overall phase also contribute to this distance. This also hints at a geometric interpretation of the dynamical phase --- it is a measure of the deviation from minimality of distance traversed in the Hilbert space. If the actual path traversed in the Hilbert space is not the one that minimizes the length, the dynamical phase is non-zero and it needs to be subtracted from the total phase in order to obtain the geometric phase. To illustrate these ideas, we consider an example loop on the Bloch sphere.  

\begin{figure*}[ht!]
\includegraphics[scale=0.6]{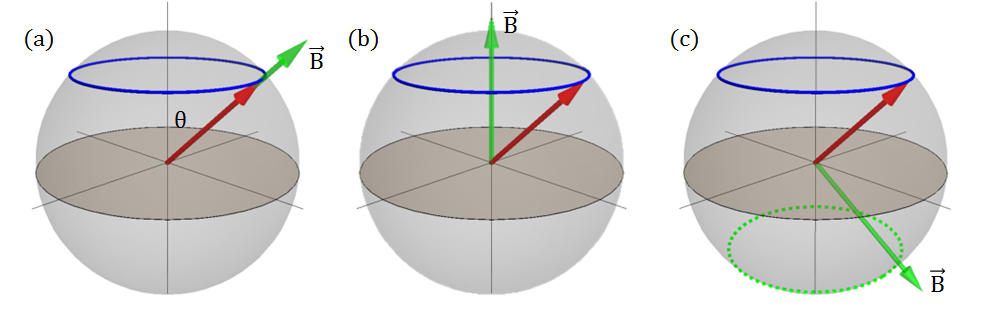}
\caption{\textbf{Dynamical Phase:} (a), (b) and (c) show three different ways of inducing a latitude in a spin-1/2 system. The magnetic field in each case is indicated by $\vec{B}$. While the geometric phase is the same for all three of them, the dynamical phase is different (see text).}\label{FIG2S}
\end{figure*}

Let us consider a latitude at $\theta$ (\figref{FIG2S}) on the Bloch sphere. Because this example is of Berry's phase (and not our geometric phase), we consider a spin-half system transported along this loop. The three obvious ways of doing this are illustrated in \figref{FIG2S}.  The familiar adiabatic change of the direction of the applied magnetic field, where the spin vector remains parallel to it throughout (this was Berry's original example) is shown in \figref{FIG2S}(a).  \figref{FIG2S}(b) shows a constant field in the $z$-direction pulsed on for a period in which the spin vector completes exactly one rotation, thereby tracing out the loop. This is the example considered in \cite{PhysRevLett.58.1593}.  \figref{FIG2S}(c) shows a magnetic field of constant magnitude, always maintained normal to the spin vector. This field transports the spin vector along the latitude, while itself traversing a different latitude. The three Hamiltonians ($H_a, H_b, H_c)$ and the corresponding times $(T_a, T_b, T_c)$ are:
\begin{equation}
\begin{split}
H_a(t) &=  \Omega \cos \theta \sigma_z + \Omega \sin \theta \cos (\omega t)\sigma_x +\Omega \sin \theta \sin (\omega t) \sigma_y \\ &\ \ \ \Omega \texttt{>>} \omega \ \& \ T_a= \frac{2\pi}{\omega}\\ 
H_b(t) &= \Omega \sigma_z: \ \ \ T_b=\frac{2\pi}{\Omega}\\
H_c(t) &=  -\Omega \sin \theta \sigma_z + \Omega \cos\theta \cos (\omega t)\sigma_x +\Omega \cos \theta \sin (\omega t) \sigma_y \\ & \ \ \ \Omega = \omega \sin \theta \ \& \ T_c= \frac{2\pi}{\omega}\\ 
\end{split}
\end{equation}
$\sigma_{x,y,z}$ are the Pauli matrices. Starting with the same initial state $|\psi\rangle$, the three Hamiltonians induce the same path on the Bloch sphere, but different paths in the Hilbert space --- they differ in the profile of the overall phase. Explicitly, the paths in the Hilbert space are,
\begin{equation}
\begin{split}
|\psi_a(t)\rangle &= e^{i \omega t \sigma_z}e^{i t \sigma_a}|\psi\rangle: \ \ \  \sigma_a= (\omega +\Omega \cos \theta)\sigma_z + \Omega \sin \theta \sigma_x\\
|\psi_b(t)\rangle &= e^{i\Omega t \sigma_z}|\psi\rangle \\  
|\psi_c(t)\rangle &= e^{i \omega t \sigma_z}e^{it \sigma_c}|\psi\rangle: \ \ \ \sigma_c= (\omega - \Omega \sin \theta)\sigma_z + \Omega \cos \theta \sigma_x\\
\end{split}
\end{equation}
The lengths of the three paths are computed using $s=\int \sqrt{\langle \dot{\psi}|\dot{\psi}\rangle}dt = \int \sqrt{\langle \psi |H^2|\psi\rangle} dt $ and the dynamical phase using $\phi_d = \int \langle \psi| H |\psi\rangle dt $ (see Ref. \cite{PhysRevLett.58.1593}). Below is a table comparing the three paths:
\begin{center}
\begin{tabular}{ |c|c|c|c|c| } 
 \hline
 Path & Path length & Total phase & Dynamical  & Geometric \\ 
 & &  &  Phase &  Phase \\ 
 \hline
 $\psi_a$ & $\frac{2\pi \Omega}{\omega}$ & $2\pi(\frac{\Omega}{\omega}-\cos\theta)$ & $\frac{2\pi \Omega}{\omega}$  & $-2\pi\cos\theta $\\ 
 $\psi_b$ & $2\pi$ & $0$ & $2\pi \cos \theta$  & $-2\pi\cos\theta $\\ 
  $\psi_c$ & $2\pi \sin \theta$ & $-2\pi\cos\theta$ & $0$  & $-2\pi\cos\theta$\\ 
 \hline
\end{tabular}
\end{center}

Clearly, $\psi_c$ has the least length among the three and in fact, among \textit{all} possible paths, because its length is equal to that of the latitude \cite{Uhlmann1995}. This is indeed the parallel transport (see ref. \cite{2016GeometricQC}). $\psi_a$ has the largest path length (because $\Omega >> \omega$) and that is reflected in the very large dynamical phase. Intuitively, dynamical phase is a unnecessary rotation of the quantum state about its own spin vector, causing the system to traverse a longer path in the Hilbert space. Such rotations have been cautiously avoided in $\psi_c$, resulting in a zero dynamical phase and minimal path length. 

The above examples illustrate two fundamental ideas regarding geometric phases --- first, that geometric phase is a purely kinematic property depending only on the geometry of the loop, regardless of the dynamics inducing the loop \cite{Mukunda-1, Mukunda-2} and second, minimization of the length is a general definition of parallel transport. Using these two ideas, we provide a mathematical definition of our geometric phase in the following section.

\section{Calculating the geometric phase}

In our geometric phase, the gauge variables are the components of the spin fluctuation tensor. The space of quantum states has a metric, known as the Fubini-Study metric($s_{FS}$), which is essentially the fidelity measure between two normalized quantum states $\psi_1, \psi_2$:
\begin{equation}
s_{FS}(\psi_1, \psi_2) = \cos^{-1}\left(|\langle \psi_1|\psi_2\rangle|\right)
\end{equation}
Hereafter, we write a loop inside the Bloch sphere parameterized by $t$ as $\vec{S}(t)$ with the parameter ranging from $0$ to $t_{final}$. Parallel transport of the ellipsoid (or the chord) along $\vec{S}(t)$ is a loop in the space of quantum states, which we may write $\psi(t)\equiv(\vec{S}, \pm \hat{u}(t))$, where, $\hat{u}(t)$ is a unit vector in space chosen such that it is always normal to $\vec{S}$ and the length of $\psi(t)$ under the Fubini-Study metric is  minimized. This condition, of minimizing the length translates to the following differential equation on $\hat{u}(t)$ \cite{Theory_Paper} :
\begin{equation}\label{eqn_u}
\frac{d}{dt}\hat{u}(t) = -\left(\frac{d}{dt}\frac{\vec{S}(t)}{|\vec{S}(t)|}\cdot \hat{u}(t)\right)\frac{\vec{S}(t)}{|\vec{S}(t)|}
\end{equation}
The parallel transport of any starting state $\psi(0)$ along $\gamma(t)$ is obtained by solving the above differential equation with the corresponding initial value of $\hat{u}(t)$.

The corresponding geometric phase, i.e., the $SO(3)$ operator $R$ is also obtained by solving a differential equation. We introduce a path $X(t)$ in $SO(3)$ which satisfies the following differential equation (we have used $\frac{\vec{S}(t)}{|\vec{S}(t)|}=\hat{v}(t)$ for simplicity here):
\begin{equation}\label{eqn_R}
\begin{split}
&\frac{d}{dt}X(t)= \left(\frac{d\hat{v}(t)}{dt}\hat{v}(t)^T-\hat{v}(t)\frac{d\hat{v}(t)}{dt}^T\right)X\\
&X(0) = 1\\
\end{split}
\end{equation}
The superscript ``$T$" indicates the transpose of a vector. The solution to this equation provides $X(t)$ and, the geometric phase of $\vec{S}(t)$ is given by $R=X(t_{final})$. Finally, the generalized solid angle is given by $\cos^{-1}(\hat{}k\cdot R\hat{k})$, where $\hat{k}$ is some vector normal to both $\hat{v}(0)$ and $\hat{v}(t_{final})$. This is the angle by which $R$ rotates a vector normal to $\hat{v}(0)$ and $\hat{v}(t_{final})$.

\begin{figure*}
\includegraphics[scale=0.6]{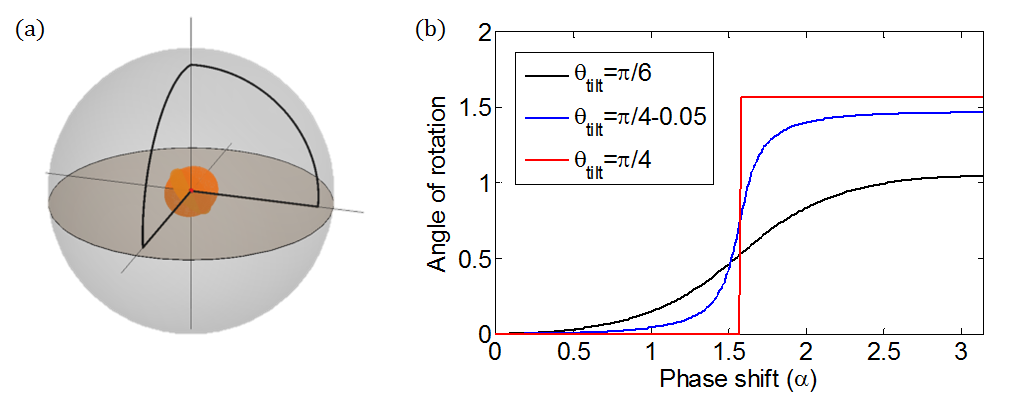}
\caption{(a) shows the loop induced in the experiment. (b) shows a plot of the angle of rotation of the ellipsoids about the spin vector, while the latter is advanced away from the center using microwaves, as a function of the phase shift $\alpha$ between the $m=0$ and $m=-1$ components. This phase is representative of the length of the spin vector, which is given by $|\vec{S}|=\sin 2\theta_{tilt}\sin \alpha$.}\label{FIG3S}
\end{figure*}

We now show how the geometric phase and generalized solid angle of the loop induced experimentally \figref{FIG3S}(a) are determined. Assuming that the loop goes out upto a radius $r$ and with the parameter $t_{final}=1$ it can be parametrized as:
\begin{equation*}
\vec{S}(t)=\begin{cases}
(4rt, 0, 0):\ \ 0\leq t\leq 1/4 \\
(  r \cos (2\pi(t-1/4)),0, r \sin (2\pi(t-1/4))):\\ \ \ 1/4 \leq t\leq 1/2\\
(r\sin\phi \sin (2\pi(t-1/2)), r\cos\phi\sin (2\pi(t-1/2)), \\ r \cos (2\pi(t-1/2))): \ \ 1/2 \leq t\leq 3/4\\
(0, 4r(1-t), 0):\ \ 3/4\leq t\leq 1 \\
\end{cases}
\end{equation*}
It is convenient to calculate $\hat{v}(t)=\frac{\vec{S}(t)}{|\vec{S}(t)|}$ :
\begin{equation*}\small
\hat{v}(t)=\begin{cases}
(1, 0, 0):\ \ 0\leq t\leq 1/4 \\
( \cos (2\pi(t-1/4)), 0,  \sin (2\pi(t-1/4))): \ \ 1/4 \leq t\leq 1/2\\
( \sin\phi \sin (2\pi(t-1/2)), \cos\phi\sin (2\pi(t-1/2)), \\  \cos (2\pi(t-1/2))): \ \ 1/2 \leq t\leq 3/4\\
(0, 1, 0):\ \ 3/4\leq t\leq 1 \\
\end{cases}
\end{equation*} 
The solution to \Eqref{eqn_R} are:
\begin{equation*}\small
X(t)=\begin{cases}
1:\ \ 0\leq t\leq 1/4 \\
R_y(-2\pi(t-1/4)): \ \ 1/4 \leq t\leq 1/2\\
R_z(\phi-\pi/2)R_x(-2\pi(t-1/2))R_z(\pi/2-\phi)R_y(-\pi/2):\\ \ \ 1/2 \leq t\leq 3/4\\
R_z(\phi)R_x(-\phi):\ \ 3/4\leq t\leq 1 \\
\end{cases}
\end{equation*}
The geometric phase is $R=X(1)=R_z(\phi)R_x(-\phi)$. Explicitly, this is a $3\times 3$ matrix:
\begin{equation}
R=\left(
\begin{array}{ccc}
\cos\phi & \sin \phi\cos\phi & 0 \\
-\sin\phi & \cos^2\phi & -\sin \phi\cos\phi \\
0& \sin \phi & \cos\phi \\
\end{array}
\right)
\end{equation}
The generalized solid angle is obtained using $\hat{k}=\hat{z}$(this is the only choice, normal to both $\hat{v}(0)$ and $\hat{v}(1)$), $\cos^{-1}(\hat{z}\cdot R\hat{z})=\phi$. 

\subsection{The dynamical phase in our loops}
In this section, we make a few remarks on the general properties of dynamical phase and show that for the loops considered in the main text, only the straight segments contribute a non-zero dynamical phase. Although our geometric phase is different from Berry's phase in that it minimizes the Fubini-Study metric as opposed to the standard Eulidean metric, it can be shown that the arcs in the induced loops contribute no dynamical phase because of the way they are induced --- by applying a field normal to the spin vector.

The straight segments, however, accumulate a non-zero dynamical phase, because they are induced by the Hamiltonian $S_z^2$, which  rotates the ellipsoid about the spin vector. \figref{FIG3S}(b) shows this rotation angle as a function of the length of the spin vector, for various starting tilt angles.

However, the straight segment contributes a zero geometric phase and therefore, it is straightforward to subtract the dynamical phase as described in the main text. \section{The generalized solid angle}
In the main text, the generalized solid angle of a loop inside the Bloch sphere was defined as the \textit{holonomy} of its diametric projection into the real projective plane ($\mathbb{RP}^2$). In this section, we address the questions of what is meant by \textit{holonomy}? Why is it equal to the solid angle of the cone generated by sweeping a diameter along the loop? and how is it a justifiable generalization of the standard solid angle? Although these questions are answered in Ref. \cite{Theory_Paper}, here we provide a more intuitive and a qualitative version of it.

\subsection{What is ``Holonomy"?}
Holonomy roughly translates to `a local quantity which captures a global property', an elementary example of which is the so called \textit{spherical excess} of a spherical triangle. While it is well known that the sum of internal angles of a spherical triangle exceeds $\pi$ by an amount known as the spherical excess,  a lesser known fact is that the spherical excess is equal to the area or the solid angle enclosed by the triangle, known as Girard's theorem.

The spherical excess is quite obviously related to parallel transports. The sum of internal angles of a spherical triangle and the sum of its external angles together sum up to $3\pi$. Therefore, the latter falls short of $2\pi$ by the spherical excess. It is easy to picture the sum of external angles --- a car driven along a spherical triangle on the earth is steered by an amount equal to the sum of the external angles \footnote{We are assuming that the car's steering rotation correctly represents the car's actual rotation. }.  While the car comes back to its original orientation, i.e., rotates effectively by $2\pi$, it's steering wheel is rotated by less than $2\pi$. This means, if the car were parallel transported, i.e., moved somehow along the spherical triangle without being steered, it would return in a different orientation, rotated by the spherical excess. 

So far, we have used only the trivial properties of a spherical triangle.  An elementary, but non-trivial property of a spherical triangle is the Girard theorem, which says that the spherical excess of a triangle is exactly equal to the enclosed solid angle. This means that the car's rotation, a local quantity, actually captures a global property --- the solid angle. Therefore, we may refer to the angle of rotation due to parallel transport as a ``holonomy". 

Naturally, when a tangent line is being parallel transported along a loop on a sphere, we expect that the distance traversed in some space is being minimized. To build an analogy with the geometric phase discussed in the previous section, tangent lines with a fixed point of tangency have one degree of freedom i.e., rotation about the point of tangency and this is the gauge variable. The full configuration space of the tangent line is a three dimensional manifold. A configuration of the tangent line is specified by three coordinates, including two of the point of tangency and one of the orientation of the tangent line. Transporting a tangent line along a loop on a sphere would correspond to a path in this configuration space. This configuration space has a nontrivial topology and is known as \textit{lens space}, $L(4,1)$. This space can be understood as a ``bundle of circles" over a sphere. That is, at each point on a sphere, a circle is attached to carry the gauge variable. This structure is known as a \textit{circle bundle} over a sphere. The rule assigning a parallel transport is known as \textit{connection form}, which, in the present case is formulated as minimization of a distance.  The solid angle of a loop on the sphere is the holonomy of the \textit{natural} connection form on this bundle. Natural here means maximally symmetric, i.e., one that does not involve an arbitrary choice (of a basis, etc) and in this case it comes from a natural metric on $L(4,1)$.  Owing to Girard's theorem, the solid angle can be \textit{defined} as the holonomy of the natural connection form.

\begin{figure*}[ht!]
\includegraphics[scale=0.64]{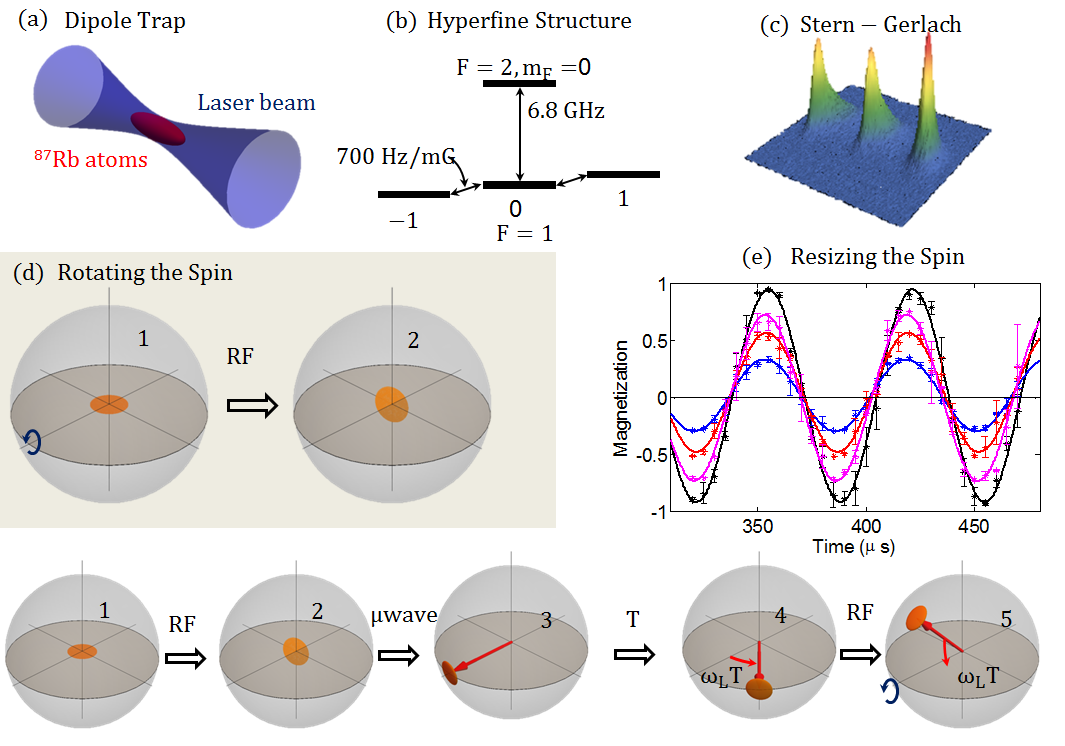}
\caption{\textbf{Our system and control operations} :  (a) shows a schematic of $^{87}$Rb atoms trapped using the dipole force in a laser field. (b) shows the hyperfine structure of $^{87}$Rb. All control operations on the $F=1$ hyperfine level are performed in the rotating frame (at Larmor frequency). (c) shows a typical Stern-Gerlach separation induced by a magnetic field gradient. (d) illustrates a controlled rotation of the quantum state. A constant magnetic field in the x-y plane in the rotating frame (i.e., a rotating magnetic field in the lab frame) rotates the quantum state. (e) shows a Ramsey sequence illustrating the control of the spin vector using RF and microwave induced transitions. In particular, the state $\psi = \frac{-1}{2}|-1\rangle + e^{i\alpha}\frac{1}{\sqrt{2}} |0\rangle + \frac{1}{2}|+1\rangle$(see text), is prepared and its spin vector is measured. The data shows a measurement of the length of the spin vector in the x-y plane, in time,  for four different values of $\alpha$. }\label{FIG4S}
\end{figure*}

\subsection{Holonomy of loops on $\mathbb{RP}^2$}

In the main text, we have shown that a non-singular loop inside the Bloch sphere can be radially projected  into the sphere and its solid angle can be defined as the holonomy of the projection. We have also shown, while singular loops can not be continuously projected to a sphere, both non-singular and singular loops can be continuously projected to the real projective plane through a diametric projection.  Therefore, the appropriate definition of generalized solid angle is the holonomy of these projections in $\mathbb{RP}^2$, provided, it agrees with the standard solid angle for the subset of non-singular loops. 

That raises the question, what is the appropriate holonomy for loops on $\mathbb{RP}^2$? Incidentally, $L(4,1)$ is also a circle bundle over $\mathbb{RP}^2$; in fact, $L(4,1)$ is also the configuration space of a unit tangent vector to $\mathbb{RP}^2$.  At each point on $\mathbb{RP}^2$, the tangent vector has a circle's worth of configurations, which form a circle in $L(4,1)$ corresponding to the point in $\mathbb{RP}^2$. This bundle also has a natural connection form that defines parallel transport of the unit tangent vector along a loop on $\mathbb{RP}^2$. The holonomy of a loop in $\mathbb{RP}^2$ is defined as the angle of rotation of a unit tangent vector when parallel transported along the loop. The corresponding connection form also comes from the same metric on $L(4,1)$ and the corresponding holonomy does agree with the standard solid angle for projection of non-singular loops \cite{Theory_Paper}. In fact, $L(4,1)$ is the only lens space that is a circle bundle over both sphere and $\mathbb{RP}^2$. 

While the generalized solid angle is a scalar, the geometric phase has been defined as an $SO(3)$ operator. Because the Bloch sphere has a singularity at the center, it is important to retain more information than just an angle of rotation. Consequently   the geometric phase, as it is defined, is the full $SO(3)$ operator.

\subsection{Holonomy of open paths in $\mathbb{RP}^2$}
 Before ending this section, we discuss the holonomy of open paths in $\mathbb{RP}^2$. Like the loop induced in the experiment, it is possible that the projection of a singular loop is an open path in $\mathbb{RP}^2$. The geometric phase, being an $SO(3)$ operator, is still well defined and represents a transformation between the tangent vectors of $\mathbb{RP}^2$ at the two endpoints of the loop. However, generalized solid angle, which is just the angle of rotation needs some clarification.
 The problem of deciding the angle between two tangent vectors at two different points on $\mathbb{RP}^2$ is analogous  to the problem of comparing the phases of two laser beams in different momentum modes and dates back to 1956 \cite{Pancharatnam1956}. The straightforward solution is to connect the two points by a geodesic and thereby close the open path. Geodesics in general have the special property that they do not accumulate any geometric phase \cite{PhysRevLett.60.2339}. 

Accordingly, the generalized solid angle is defined as follows: if $R$ is the geometric phase of a loop whose projection is open in $\mathbb{RP}^2$ and $d_1$ and $d_2$ are its endpoints (i.e, the diameters to a sphere representing the initial and final points on $\mathbb{RP}^2$), the generalized solid angle is
\begin{equation}
\Omega = \cos^{-1}(\hat{k}\cdot R\hat{k})
\end{equation}  
for some unit vector $\hat{k}$ which is normal to both $d_1$ and $d_2$. If $d_1=d_2$, i.e., if the path is closed in $\mathbb{RP}^2$, $\Omega$ is simply the angle of rotation of $R$. If $d_1\neq d_2$, the above expression provides the holonomy of the loop obtained by closing the path using a geodesic in $\mathbb{RP}^2$.  

Note that in cases where the projected path in $\mathbb{RP}^2$ is open, the starting point of the loop can not be arbitrarily chosen. The experimentally chosen loops, for instance have to start and end at the center of the Bloch sphere -- no other point can be chosen as the starting point in order to be able to lift the loop, i.e., have a continuous diametric projection. In other words, the loops chosen experimentally are \textit{liftable} only when they start and end at the center. For further details and a rigorous discussion of a \textit{liftability criterion}, see \cite{Theory_Paper}.  
\section{Experimental quantum control}

In this section, we summarize how the spin vector of ultracold $^{87}$Rb atoms is experimentally controlled. The internal state of the atoms within the $F=1$ hyperfine level (see \figref{FIG4S}(b)) can be controlled using microwaves and magnetic fields rotating at radio frequency. An arbitrary control is brought about by a combination of rotation and resizing of the spin vector inside the Bloch sphere. To suppress any noise in the magnetic field, we operate the system at a fixed applied ambient field of $20$ mG in the $z$-direction. The linear Zeeman splitting of the $m_F=0,\pm 1$  states is $700$ Hz/mG and therefore, the system undergoes a constant Larmor precession at $14$ kHz about the $z$-axis. A rotation of the spin vector about an arbitrary axis within the $x-y$ plane by an arbitrary angle can be performed using a magnetic field rotating in the x-y plane at $14$ kHz. This is engineered by two coils placed to produce magnetic fields in two orthogonal directions, driven out of phase. Although a single coil would be sufficient under the rotating wave approximation (RWA), at the required frequencies, fast RF rotations would see a breakdown of the RWA \cite{PhysRev.57.522, PhysRevA.88.054301}. A rotation about an axis in the $x-y$ plane making an angle $\xi$ with the $x$-axis, i.e., $\hat{x}\cos \xi +\hat{y}\sin \xi$ by an angle $\eta$ can be brought about by an RF pulse of pulse length $\eta$ and starting phase $\xi$. An arbitrary SO(3) operator can be constructed by composing such rotations. 

 The length of the spin vector can be controlled by a detuned $\pi$ transition between $|F=1, m_F=0\rangle$ and $|F=2, m_F=0\rangle$ levels induced by microwaves (i.e., clock transition). The energy gap between these two levels is $\Delta = 6.8 $ GHz. A $\pi-$ transition at a detuning of $\delta$ changes the phase of $|F=1, m_F=0\rangle$ relative to $|F=1, m_F=\pm 1\rangle$ by $\alpha = \pi \left(1-\frac{\delta}{\sqrt{\Omega^2 +\delta^2}}\right)$, where $\Omega$ is the rate of the microwave transition at zero detuning. For instance, the state  $\psi = \frac{-1}{2}|-1\rangle + \frac{1}{\sqrt{2}} |0\rangle + \frac{1}{2}|+1\rangle$ is transformed to $\psi' = \frac{-1}{2}|-1\rangle + e^{i\alpha}\frac{1}{\sqrt{2}} |0\rangle + \frac{1}{2}|+1\rangle$ . The former has a spin vector $\vec{S}=(0,0,0)^T$ while the latter has, $\vec{S'}= (\sin \alpha, 0, 0)^T$(see \figref{FIG4S}(e)). This technique can be viewed as a dressed Hamiltonian, $H=S_z^2$. 

\bibliography{GeometricPhasePaper}{}
\bibliographystyle{ieeetr}
\end{document}